\documentclass[prl,twocolumn,showpacs,superscriptaddress,10pt]{revtex4-1} %
\usepackage{amsmath,amssymb,graphicx}
\usepackage{txfonts}

\begin{document}
\title{Polarization control of quantum dot emission  by chiral photonic crystal slabs}

\author{Sergey V. Lobanov}\email{Corresponding author: lobanov@physics.msu.ru}
\affiliation{Skolkovo Institute of Science and Technology, Novaya Street 100, Skolkovo 143025, Russia}
\affiliation{M. V. Lomonosov Moscow State University, Leninskie Gory 1, Moscow 119991, Russia}

\author{Thomas Weiss}
\affiliation{4$^\mathrm{th}$ Physics Institute and Research Centers Scope, University of Stuttgart, D-70550 Stuttgart, Germany}

\author{Nikolay A. Gippius}
\affiliation{Skolkovo Institute of Science and Technology, Novaya Street 100, Skolkovo 143025, Russia}
\affiliation{A. M. Prokhorov General Physics Institute, Russian Academy of Sciences, Vavilova Street 38, Moscow 119991, Russia}

\author{Sergei G. Tikhodeev}
\affiliation{A. M. Prokhorov General Physics Institute, Russian Academy of Sciences, Vavilova Street 38, Moscow 119991, Russia}
\affiliation{M. V. Lomonosov Moscow State University, Leninskie Gory 1, Moscow 119991, Russia}

\author{Vladimir D. Kulakovskii}
\affiliation{Institute of Solid State Physics, Russian Academy of Science, Chernogolovka 142432, Russia}

\author{Kuniaki Konishi}
\affiliation{Institute for Photon Science and Technology, The University of Tokyo, Tokyo 113-0033, Japan}

\author{Makoto Kuwata-Gonokami}
\affiliation{Photon Science Center, The University of Tokyo, Tokyo 113-8656, Japan}
\affiliation{Department of Physics, The University of Tokyo, Tokyo 113-0033, Japan}

\begin{abstract}
We investigate theoretically the polarization properties of the quantum dot's optical emission from chiral photonic crystal structures made of achiral materials in the absence of external magnetic field at room temperature.
The mirror symmetry of the local electromagnetic field is broken in this system due to the decreased symmetry of the chiral modulated layer.
As a result, the radiation of randomly polarized quantum dots normal to the structure becomes partially circularly polarized.
The sign and degree of circular polarization are determined by the geometry of the chiral modulated structure and depend on the radiation frequency.
A degree of circular polarization up to 99\% can be achieved for randomly distributed quantum dots, and can be close to 100\% for some single quantum dots.
\end{abstract}

\maketitle


The possibility to control the circular polarization state of radiation from quantum emitters has drawn attention of researchers in recent years.
The reason for this are the various important applications in spin-optoelectronics, quantum
information technology, chiral synthesis and sensing, etc..
There are several methods to generate circularly polarized emission.
The first one is to use a quarter-wave plate.
The second possibility is based on the electrical injection of spin-polarized carriers into the active region~\cite{Fiederling1999,Ohno1999,Rashba2000,Jiang2005,Holub2007,Bhattacharya2011}.
The third possibility is to modify the local electromagnetic field, for example, using chiral liquid crystals~\cite{Stegemeyer1979,Schmidtke2003,Woon2005}.
Another possibility is the combination of the second and the third ways.
It has been suggested to use external magnetic field to split opposite spin states and microcavity to enhance the emission for some state~\cite{Ren2012}.
Modification of local electromagnetic field can also be achieved by chiral photonic crystals~\cite{Konishi2011}.
This method has considerable advantages in comparison to the others, for example, the controllability of the properties, the small structure size (a quater-wave plate is thicker), simple operation, and compatibility of semiconductor fabrication process.
The circular polarization degree (CPD) reported in~\cite{Konishi2011} was about 26\% maximum.
Recently, the CPD of QDs emission as high as 81\%  was experimentally achieved from a  planar semiconductor microcavity with
chiral half-etched top mirror~\cite{Maksimov2014},  but this structure is far more complex.
Therefore finding an optimized simpler design for the type of structure used in~\cite{Konishi2011} with a larger CPD can be clearly beneficial.


In this paper, it is shown that a CPD close to 100\% can be achieved by a simple optimization of only the chiral layer of the structure of~\cite{Konishi2011}.
In addition, the calculation in~\cite{Konishi2011} was done only for a single photon energy and emission normal to the system plane. Here we show that the calculated optical emission of quantum dots (QDs) from the structure of~\cite{Konishi2011} with gammadions is
in a good qualitative agreement with the experimental results. However some discrepancies appear, which are analyzed here.

\begin{figure}
\includegraphics{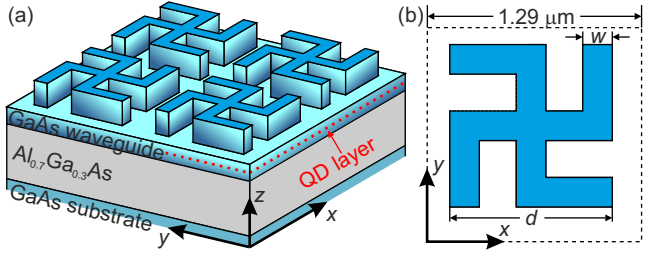}
\caption{
(a) Three-dimensional side view of the chiral nanostructure.
It consists of three homogeneous layers
(GaAs waveguide, Al$_{0.7}$Ga$_{0.3}$As underlayer and GaAs substrate) and one horizontally modulated layer (chiral photonic crystal).
Dotted red lines indicate a plane with randomly distributed QDs.
(b) Top view of the chiral nanostructure with GaAs gammadions. One unit cell is shown. Blue (white) color indicates GaAs (air).
}\label{Structure}
\end{figure}

The gammadion nanostructures consist of one chiral photonic crystal layer, three homogeneous layers, and a plane with randomly distributed QDs (Fig.~\ref{Structure}a).
The plane with QDs is located 150~nm above the bottom of 260~nm thick GaAs  wave\-guide (permittivity $\varepsilon_{\mathrm{GaAs}}\approx12.2$).
The GaAs wave\-guide layer is surrounded by two lower effective dielectric permittivity layers.
The lower layer  is 1~$\mu$m thick and consists of Al$_{0.7}$Ga$_{0.3}$As ($\varepsilon_{\mathrm{Al}_{0.7}\mathrm{Ga}_{0.3}\mathrm{As}}\approx9.3$).
The upper layer (460~nm thick)  is a square lattice (period is $p=1.29~\mu$m) of GaAs gammadions.
One unit cell of the chiral photonic crystal is shown in Fig.~\ref{Structure}b.
Each gammadion is homogeneous along $z$-axis and is characterized by 
size $d$ and arm width $w$.
The whole system is placed on a GaAs substrate.
In the numerical calculations, the substrate and the superstrate are assumed to be infinite half-spaces.

The spontaneous emission of a QD is not an intrinsic property but depends on the environment~\cite{Purcell1946,Bykov1972}.
The environment can modify all QD's optical characteristics such as the QD's excited-state lifetime, far-field directional pattern and polarization of emitted light.
If a QD is placed in a structure supporting optical modes, the QD's optical characteristics can have resonance behavior.
In this paper, we consider the structures supporting quasi-waveguide modes, which appear from the waveguide modes coupled to photons in air via the photonic crystal layer~\cite{Tikhodeev2002}.
The coupling strength is different for left ($\sigma^-$) and right ($\sigma^+$) circularly polarized photons
because of the coupling layer's symmetry.
For specificity, the left and right circular polarizations are defined from the point of view of observer.
The QDs in the structure of interest excite the quasi-waveguide modes, that leak outside the waveguide along $z$-axis with different intensities in left and right polarizations.
In order to characterize the emission properties along $z$-axis, we compute the intensities in left and right polarizations ($I^-$ and $I^+$)  and calculate the CPD $P =  (I^--I^+)/(I^-+I^+)$.

\begin{figure}[b]
\includegraphics{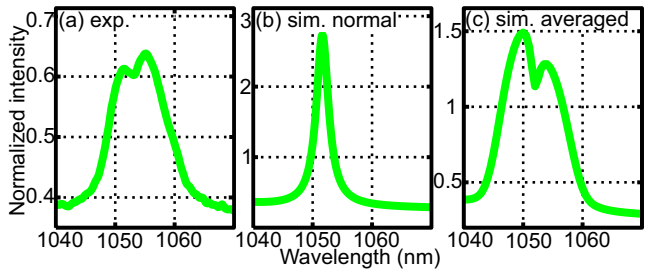}
\caption{
Left circularly polarized component of the QD's emission spectra for the structure with gammadion parameters ($w$, $d$) = (146~nm, 1074~nm).
Panels (a), (b), and (c) correspond to the measured intensity published in~\cite{Konishi2011}, calculated intensity normal to the structure, and calculated intensity averaged over a small solid angle.
See the explanation of spectra normalization and averaging in the text.
}\label{Spectra}
\end{figure}

\begin{figure}[t]
\includegraphics{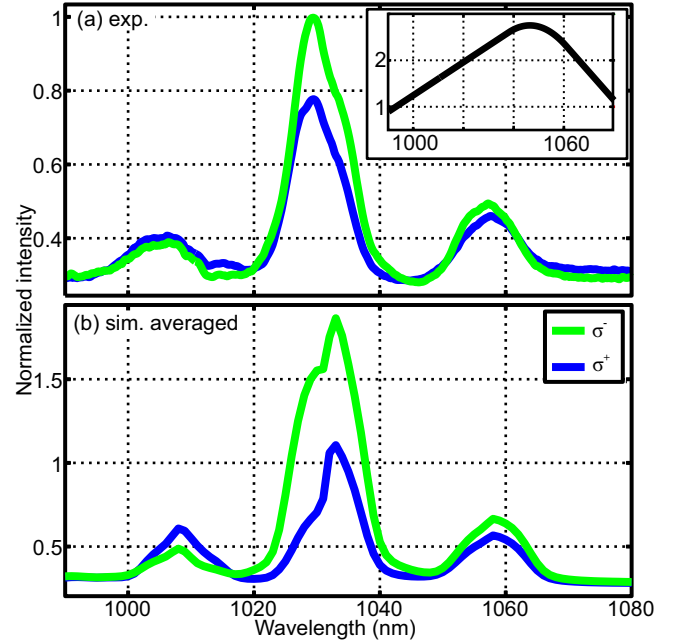}
\caption{
Measured (a)~\cite{Konishi2011} and calculated (b) left ($\sigma^-$) and right ($\sigma^+$) circularly polarized components of the QD's emission spectra
for the structure with gammadion parameters ($w$, $d$) = (177~nm, 1075~nm).
The calculated intensities in panel (b) are averaged over a small solid angle, as explained in the text. The inset in panel (a) shows the experimental QD's emission spectra
prior to chiral layer fabrication, which is same in left and right circular polarizations.
}\label{Spectra_177}
\end{figure}

The computation of the QD's radiation can be performed using an oscillating point dipole model.
In this model, a finite-size QD in the point ${\bf r}_0$ is replaced by an oscillating point current ${\bf j}({\bf r},t)={\bf j}_0\delta({\bf r}-{\bf r}_0)\exp(-i\omega t)$ with fixed amplitude $\bf{j}_0$ and frequency $\omega$ (so-called weak coupling limit).
This system is described by Maxwell's equations and the far-field emission can be computed using the scattering matrix treatment~\cite{Tikhodeev2002,Whittaker1999,Weiss2009,Lobanov2012}.
As the QDs in the structure of interest are randomly distributed and non-polarized, the radiation intensity of the oscillating point dipole must be averaged over different dipole positions ${\bf r}_0$ and directions of dipole moment.
We assume also that the oscillating point dipole can oscillate only along the $xy$-plane because of the QD's shape.

The experimental photoluminescence spectrum of the QD's radiation along $z$ axis in the left circular polarization for the structure with gammadion parameters ($w$, $d$) = (146~nm, 1074~nm) is shown in Fig.~\ref{Spectra}a (see also~\cite{Konishi2011}).
In this figure and Fig.~\ref{Spectra_177}a, the experimental emission intensity is normalized to the function shown in the inset of Fig.~\ref{Spectra_177}a and a constant such that non-resonant experimental and theoretical intensities are close to each other.
The corresponding calculated spectrum of the QD's emission normal to the structure is shown in Fig.~\ref{Spectra}b.
In these and following figures, the calculated intensity is normalized to the maximum radiation intensity of the equivalent QD in free space.
One resonance (with wavelengths $\lambda\approx 1052$ nm) is shown in the QD's emission spectra.
The position of this resonance for the experimental and calculated spectra are almost the same, but the width of the experimental resonances is several times broader than the theoretical and the amplitude is several times smaller.
It may be due to a number of reasons.
The first probable reason for this is that there are many defects in the real structure, leading to additional light scattering.
As a result, the resonances in the QD's emission spectra for the real structure should be smaller and broader than for the ideal one.
In addition, this mechanism mixes emission from different directions.
In particular, emission spectra to inclined directions make a small contribution to the experimental emission spectrum to the top.
The defects are not explicitly taken into account here, as it is rather difficult to include this contribution in numerical calculations.

\begin{figure}
\includegraphics{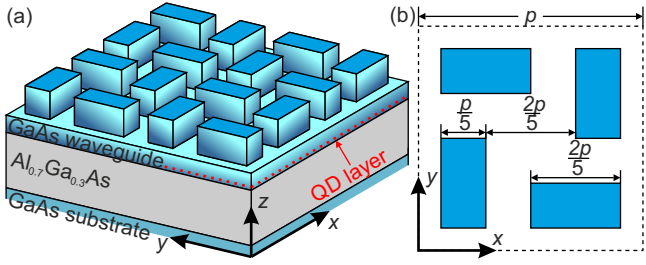}
\caption{
(a) Three-dimensional side view and (b) top view of the optimized chiral nanostructure.
}\label{Chiral}
\end{figure}

Another probable reason is a finite aperture value in the experimental setup (NA~$\sim0.03$).
The resonance frequency depends on the angle of observation.
The stronger this dependence and the larger the aperture, the smaller and broader are the resonances.
To take into account the finite aperture, we average the intensity $I(\lambda, \theta, \phi)$ over different directions with the weight $f(\theta)= 1 / ( 1 + \exp((\theta - \theta_0)/\Delta\theta))$,
where $\theta_0 = 0.025^\mathrm{rad}$ and $\Delta\theta = 0.003^\mathrm{rad}$:
\begin{align}
I_0 (\lambda) = \frac{1}{N}\int_0^{\pi}\mathrm{d}\theta \sin\theta f(\theta) \int_0^{2\pi} \mathrm{d}\phi I(\lambda, \theta, \phi). \label{Aver}
\end{align}
Here, $I_0 (\lambda)$ is averaged intensity, $\theta$ and $\phi$ are polar and azimuth angles, and $N = 2\pi\int_0^{\pi}\mathrm{d}\theta \sin\theta f(\theta)$.
The corresponding spectrum is shown in Fig.~\ref{Spectra}c.
This spectrum looks more similar to the measured one than the non-averaged emission spectrum.
It has the same double-resonance shape and the same width as the experimental resonance.
Nevertheless, its amplitude is still several times bigger than the experimental one, indicating the presence of defects and other reasons for the observed discrepancy.

One more probable reason for the difference between the calculated and experimental spectra is a very strong dependence of emission spectra on the structure's parameters.
For instance, changing the arm width $w$ for the structure with $w=146$~nm by only 1\% leads to a change in the amplitude shown in Fig.~\ref{Spectra}b by 12\%.

Figure~\ref{Spectra_177}a reproduces the emission spectra of Ref.~\cite{Konishi2011} for the structure with gammadion parameters ($w$, $d$) = (177~nm, 1075~nm).
The calculated intensities averaged over small solid angle with slightly modified structural parameters (the etching thickness is 19~nm smaller, the size of the gammadion $d$ is 11~nm smaller, the outer arm is 3~nm wider and 2~nm shorter, the inner $x$-directed arm is 14~nm narrower, and the inner $y$-directed arm is 4~nm broader) are shown in Fig.~\ref{Spectra_177}b.
One can observe a good overall agreement between measured and calculated spectra in Fig.~\ref{Spectra_177}.

\begin{figure}
\includegraphics{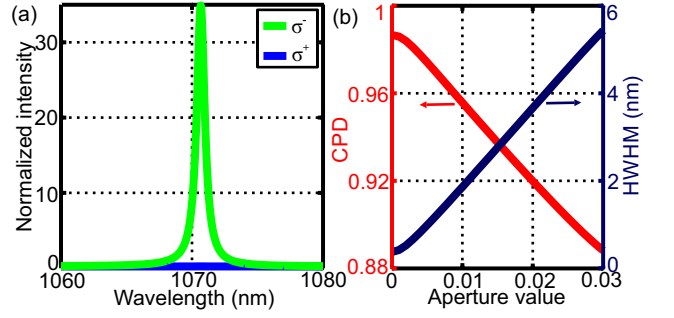}
\caption{
(a) Calculated left ($\sigma^-$) and right ($\sigma^+$) circularly polarized components of the QD's emission spectra for the optimized chiral nanostructure in Fig.~\ref{Chiral}.
(b) Calculated dependencies of CPD at resonance wavelength and HWHM of the resonance on aperture value.
}\label{Optimized_structure_Spectra}
\end{figure}


\begin{figure*}
\includegraphics{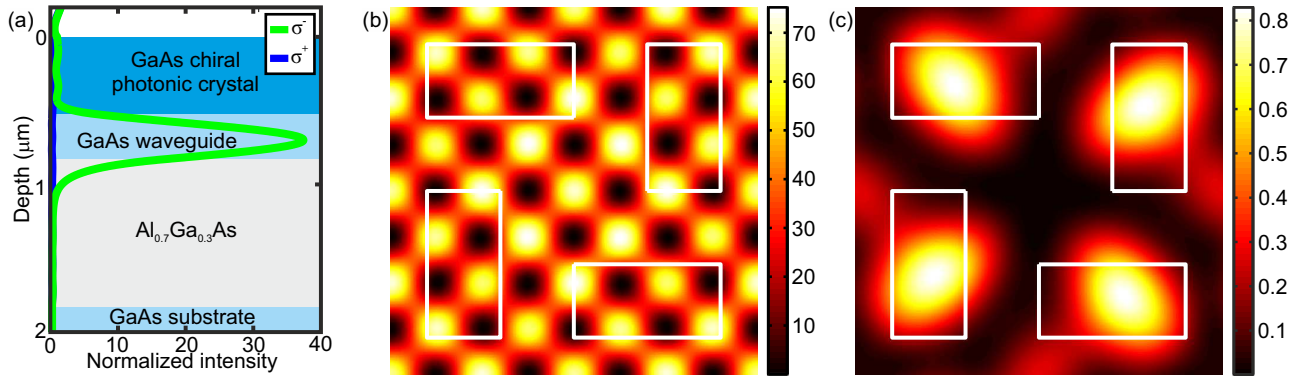}
\caption{
(a) Calculated dependencies of left ($\sigma^-$) and right ($\sigma^+$) circularly polarized components of the far-field emission on the vertical position of the layer with randomly distributed QDs.
The blue, light blue, and gray rectangles mark the chiral photonic crystal layer, GaAs waveguide, and Al$_{0.7}$Ga$_{0.3}$As layer, respectively.
(b,c) Calculated dependencies of (b) left and (c) right circularly polarized components of the single QD's far-field emission along $z$~direction on its horizontal position (depth is 740 nm).
One unit cell (period is $p=920$~nm) is shown in all pictures.
White rectangles mark horizontal positions of chiral photonic crystal elements.
}\label{XYZ_dependence}
\end{figure*}

So far, we have verified experimental results of~\cite{Konishi2011}. In the following, we provide an alternative design for the photonic crystal slab, which leads to nearly 100\% CPD in our numerical calculations and is even more simple than the gammadions.
The optimized structure consists of the same layers as the previous design, but the gammadions are replaced by four rectangles, each rotated by $90^{\mathrm{o}}$ with respect to its nearest neighbors. A schematic of the structure is depicted in Fig.~\ref{Chiral}.
The thickness of the AlGaAs layer is the same as in the previous design, while the thicknesses of the chiral photonic crystal layer and the waveguide are 520 nm and 310 nm, respectively.
The period of the chiral photonic crystal is $p = 920$ nm and its unit cell is shown in Fig.~\ref{Chiral}b.
The distance between rectangles is $p/5$, their sides are $p/5$ and $2p/5$.
The plane with randomly distributed QDs is located at 90~nm above the waveguide's bottom.

The calculated spectra of the QD's emission for the optimized nanostructure are shown in Fig.~\ref{Optimized_structure_Spectra}a.
One strong and narrow resonance for left polarized photons occurs at a wavelength of 1071 nm.
In contrast, emission in the right polarization does not provide any resonance behavior at this wavelength.
The CPD appears to be around~$+99\%$ at resonance wavelength.
The influence of a finite aperture can be estimated using the same formula~\ref{Aver} as before and linear dependence of $\theta_0$ and $\Delta \theta$ on aperture value.
Figure~\ref{Optimized_structure_Spectra}b shows that the CPD decreases almost linearly with increasing aperture
while the resonance peak half width at half maximum (HWHM) increases almost linearly.
It can be seen that CPD larger than 95\% is expected in this structure within aperture less than 0.01.

Emission properties depend also on the vertical position of the QDs.
Such calculated dependence of left ($\sigma^-$) and right ($\sigma^+$) circularly polarized components of the emission are shown in Fig.~\ref{XYZ_dependence}a.
The left circularly polarized component of the emission reaches its maximum for QDs located inside the waveguide and decreases rapidly outside.
It proves the fact that the QDs couple with a quasi-waveguide mode which can leak along the positive $z$ axis with left circular polarization.
The behavior of the right circularly polarized component of the emission is not the same due to the broken mirror symmetry and optimized parameters of the chiral photonic crystal layer.

The dependence of single QD's emission on its horizontal position is shown in Figs.~\ref{XYZ_dependence}b-\ref{XYZ_dependence}c for left and right circular polarized output to the superstrate.
The figures show that QDs excite a quasi-waveguide mode originating from a standing waveguide mode~\cite{Tikhodeev2002}, for which the electric field is proportional to $\cos(4\pi x / p)\cos(4\pi y / p)$.
The maximum value of left polarized emission exceeds the maximum value of right polarized emission by a factor of one hundred.
It can be seen also that right polarized emission is about zero for some QD's position while left polarized emission is large. That means the CPD is close to $+100\%$ for these points.

In conclusion, we theoretically investigated emission of randomly distributed quantum dots from chiral photonic crystal structures.
The calculated results for gammadion nanostructures agree with the experimental ones~\cite{Konishi2011}.
We propose an optimized chiral nanostructure, which enhances left polarized emission, leading to a CPD around 99\%.
The control of quantum dot's position leads to increasing of CPD up to 100\%.

This work was supported in part by RFBR,
Russian Federation Presidential Council for Grants,
the Dynasty Foundation, EU project SPANGL4Q,
the Photon Frontier Network Program,
KAKENHI(20104002), Project for Developing Innovation Systems of
the Ministry of Education, Culture, Sports, Science and Technology (MEXT), Japan, and
by JSPS through its FIRST Program.


\begin{thebibliography}{19}%
\makeatletter
\providecommand \@ifxundefined [1]{%
 \@ifx{#1\undefined}
}%
\providecommand \@ifnum [1]{%
 \ifnum #1\expandafter \@firstoftwo
 \else \expandafter \@secondoftwo
 \fi
}%
\providecommand \@ifx [1]{%
 \ifx #1\expandafter \@firstoftwo
 \else \expandafter \@secondoftwo
 \fi
}%
\providecommand \natexlab [1]{#1}%
\providecommand \enquote  [1]{``#1''}%
\providecommand \bibnamefont  [1]{#1}%
\providecommand \bibfnamefont [1]{#1}%
\providecommand \citenamefont [1]{#1}%
\providecommand \href@noop [0]{\@secondoftwo}%
\providecommand \href [0]{\begingroup \@sanitize@url \@href}%
\providecommand \@href[1]{\@@startlink{#1}\@@href}%
\providecommand \@@href[1]{\endgroup#1\@@endlink}%
\providecommand \@sanitize@url [0]{\catcode `\\12\catcode `\$12\catcode
  `\&12\catcode `\#12\catcode `\^12\catcode `\_12\catcode `\%12\relax}%
\providecommand \@@startlink[1]{}%
\providecommand \@@endlink[0]{}%
\providecommand \url  [0]{\begingroup\@sanitize@url \@url }%
\providecommand \@url [1]{\endgroup\@href {#1}{\urlprefix }}%
\providecommand \urlprefix  [0]{URL }%
\providecommand \Eprint [0]{\href }%
\providecommand \doibase [0]{http://dx.doi.org/}%
\providecommand \selectlanguage [0]{\@gobble}%
\providecommand \bibinfo  [0]{\@secondoftwo}%
\providecommand \bibfield  [0]{\@secondoftwo}%
\providecommand \translation [1]{[#1]}%
\providecommand \BibitemOpen [0]{}%
\providecommand \bibitemStop [0]{}%
\providecommand \bibitemNoStop [0]{.\EOS\space}%
\providecommand \EOS [0]{\spacefactor3000\relax}%
\providecommand \BibitemShut  [1]{\csname bibitem#1\endcsname}%
\let\auto@bib@innerbib\@empty
\bibitem [{\citenamefont {Fiederling}\ \emph {et~al.}(1999)\citenamefont
  {Fiederling}, \citenamefont {Keim}, \citenamefont {Reuscher}, \citenamefont
  {Ossau}, \citenamefont {Schmidt}, \citenamefont {Waag},\ and\ \citenamefont
  {Molenkamp}}]{Fiederling1999}%
  \BibitemOpen
  \bibfield  {author} {\bibinfo {author} {\bibfnamefont {R.}~\bibnamefont
  {Fiederling}}, \bibinfo {author} {\bibfnamefont {M.}~\bibnamefont {Keim}},
  \bibinfo {author} {\bibfnamefont {G.}~\bibnamefont {Reuscher}}, \bibinfo
  {author} {\bibfnamefont {W.}~\bibnamefont {Ossau}}, \bibinfo {author}
  {\bibfnamefont {G.}~\bibnamefont {Schmidt}}, \bibinfo {author} {\bibfnamefont
  {A.}~\bibnamefont {Waag}}, \ and\ \bibinfo {author} {\bibfnamefont {L.~W.}\
  \bibnamefont {Molenkamp}},\ }
  \bibfield {title} {Injection and detection of a spin-polarized current in a light-emitting diode}, \
  \href
  {http://www.nature.com/nature/journal/v402/n6763/abs/402787a0.html}
  {\bibfield  {journal} {\bibinfo  {journal} {Nature}\ }\textbf {\bibinfo
  {volume} {402}},\ \bibinfo {pages} {787} (\bibinfo {year}
  {1999})}\BibitemShut {NoStop}%
\bibitem [{\citenamefont {Ohno}\ \emph {et~al.}(1999)\citenamefont {Ohno},
  \citenamefont {Young}, \citenamefont {Beschoten}, \citenamefont {Matsukura},
  \citenamefont {Ohno},\ and\ \citenamefont {Awschalom}}]{Ohno1999}%
  \BibitemOpen
  \bibfield  {author} {\bibinfo {author} {\bibfnamefont {Y.}~\bibnamefont
  {Ohno}}, \bibinfo {author} {\bibfnamefont {D.~K.}\ \bibnamefont {Young}},
  \bibinfo {author} {\bibfnamefont {B.}~\bibnamefont {Beschoten}}, \bibinfo
  {author} {\bibfnamefont {F.}~\bibnamefont {Matsukura}}, \bibinfo {author}
  {\bibfnamefont {H.}~\bibnamefont {Ohno}}, \ and\ \bibinfo {author}
  {\bibfnamefont {D.~D.}\ \bibnamefont {Awschalom}},\ }
  \bibfield {title} {Electrical spin injection in a ferromagnetic semiconductor heterostructure}, \
  \href
  {http://www.nature.com/nature/journal/v402/n6763/abs/402790a0.html}
  {\bibfield  {journal} {\bibinfo  {journal} {Nature}\ }\textbf {\bibinfo
  {volume} {402}},\ \bibinfo {pages} {790} (\bibinfo {year}
  {1999})}\BibitemShut {NoStop}%
\bibitem [{\citenamefont {Rashba}(2000)}]{Rashba2000}%
  \BibitemOpen
  \bibfield  {author} {\bibinfo {author} {\bibfnamefont {E.~I.}\ \bibnamefont
  {Rashba}},\ }
  \bibfield {title} {Theory of electrical spin injection: Tunnel contacts as a solution of the conductivity mismatch problem}, \
  \href {http://link.aps.org/doi/10.1103/PhysRevB.62.R16267}
  {\bibfield  {journal} {\bibinfo  {journal} {Physical Review B}\ }\textbf
  {\bibinfo {volume} {62}},\ \bibinfo {pages} {R16267} (\bibinfo {year}
  {2000})}\BibitemShut {NoStop}%
\bibitem [{\citenamefont {Jiang}\ \emph {et~al.}(2005)\citenamefont {Jiang},
  \citenamefont {Wang}, \citenamefont {Shelby}, \citenamefont {Macfarlane},
  \citenamefont {Bank}, \citenamefont {Harris},\ and\ \citenamefont
  {Parkin}}]{Jiang2005}%
  \BibitemOpen
  \bibfield  {author} {\bibinfo {author} {\bibfnamefont {X.}~\bibnamefont
  {Jiang}}, \bibinfo {author} {\bibfnamefont {R.}~\bibnamefont {Wang}},
  \bibinfo {author} {\bibfnamefont {R.~M.}\ \bibnamefont {Shelby}}, \bibinfo
  {author} {\bibfnamefont {R.~M.}\ \bibnamefont {Macfarlane}}, \bibinfo
  {author} {\bibfnamefont {S.~R.}\ \bibnamefont {Bank}}, \bibinfo {author}
  {\bibfnamefont {J.~S.}\ \bibnamefont {Harris}}, \ and\ \bibinfo {author}
  {\bibfnamefont {S.~S.~P.}\ \bibnamefont {Parkin}},\ }
  \bibfield {title} {Highly Spin-Polarized Room-Temperature Tunnel Injector for Semiconductor Spintronics using MgO(100)}, \
  \href
  {http://link.aps.org/doi/10.1103/PhysRevLett.94.056601} {\bibfield  {journal}
  {\bibinfo  {journal} {Physical Review Letters}\ }\textbf {\bibinfo {volume}
  {94}},\ \bibinfo {pages} {56601} (\bibinfo {year} {2005})}\BibitemShut
  {NoStop}%
\bibitem [{\citenamefont {Holub}\ and\ \citenamefont
  {Bhattacharya}(2007)}]{Holub2007}%
  \BibitemOpen
  \bibfield  {author} {\bibinfo {author} {\bibfnamefont {M.}~\bibnamefont
  {Holub}}\ and\ \bibinfo {author} {\bibfnamefont {P.}~\bibnamefont
  {Bhattacharya}},\ }
  \bibfield {title} {Spin-polarized light-emitting diodes and lasers}, \
  \href
  {http://lib.semi.ac.cn:8080/tsh/dzzy/wsqk/IOP/J-Phys-D/40-R179.pdf}
  {\bibfield  {journal} {\bibinfo  {journal} {J. Phys. D: Appl. Phys.}\
  }\textbf {\bibinfo {volume} {40}},\ \bibinfo {pages} {R179} (\bibinfo {year}
  {2007})}\BibitemShut {NoStop}%
\bibitem [{\citenamefont {Bhattacharya}\ \emph {et~al.}(2011)\citenamefont
  {Bhattacharya}, \citenamefont {Basu}, \citenamefont {Das},\ and\
  \citenamefont {Saha}}]{Bhattacharya2011}%
  \BibitemOpen
  \bibfield  {author} {\bibinfo {author} {\bibfnamefont {P.}~\bibnamefont
  {Bhattacharya}}, \bibinfo {author} {\bibfnamefont {D.}~\bibnamefont {Basu}},
  \bibinfo {author} {\bibfnamefont {A.}~\bibnamefont {Das}}, \ and\ \bibinfo
  {author} {\bibfnamefont {D.}~\bibnamefont {Saha}},\ }
  \bibfield {title} {Quantum dot polarized light sources}, \
  \href {\doibase
  10.1088/0268-1242/26/1/014002} {\bibfield  {journal} {\bibinfo  {journal}
  {Semiconductor Science and Technology}\ }\textbf {\bibinfo {volume} {26}},\
  \bibinfo {pages} {014002} (\bibinfo {year} {2011})}\BibitemShut {NoStop}%
\bibitem [{\citenamefont {Stegemeyer}\ \emph {et~al.}(1979)\citenamefont
  {Stegemeyer}, \citenamefont {Stille},\ and\ \citenamefont
  {Pollmann}}]{Stegemeyer1979}%
  \BibitemOpen
  \bibfield  {author} {\bibinfo {author} {\bibfnamefont {H.}~\bibnamefont
  {Stegemeyer}}, \bibinfo {author} {\bibfnamefont {W.}~\bibnamefont {Stille}},
  \ and\ \bibinfo {author} {\bibfnamefont {P.}~\bibnamefont {Pollmann}},\
  }
  \bibfield {title} {Circular Fluorescence Polarization of Achiral Molecules in Cholesteric Liquid Crystals}, \
  \href
  {http://scholar.google.com/scholar?hl=en&btnG=Search&q=intitle:Circular+Fluorescence+Polarization+of+Achiral+Molecules+in+Cholesteric+Liquid+Crystals#0}
  {\bibfield  {journal} {\bibinfo  {journal} {Israel J Chem}\ }\textbf
  {\bibinfo {volume} {18}},\ \bibinfo {pages} {312} (\bibinfo {year}
  {1979})}\BibitemShut {NoStop}%
\bibitem [{\citenamefont {Schmidtke}\ and\ \citenamefont
  {Stille}(2003)}]{Schmidtke2003}%
  \BibitemOpen
  \bibfield  {author} {\bibinfo {author} {\bibfnamefont {J.}~\bibnamefont
  {Schmidtke}}\ and\ \bibinfo {author} {\bibfnamefont {W.}~\bibnamefont
  {Stille}},\ }
  \bibfield {title} {Fluorescence of a dye-doped cholesteric liquid crystal film in the region of the stop band: theory and experiment}, \
  \href {\doibase 10.1140/epjb/e2003-00022-x} {\bibfield
  {journal} {\bibinfo  {journal} {The European Physical Journal B - Condensed
  Matter and Complex Systems}\ }\textbf {\bibinfo {volume} {31}},\ \bibinfo
  {pages} {179} (\bibinfo {year} {2003})}\BibitemShut {NoStop}%
\bibitem [{\citenamefont {Woon}\ \emph {et~al.}(2005)\citenamefont {Woon},
  \citenamefont {O’Neill}, \citenamefont {Richards}, \citenamefont {Aldred},\
  and\ \citenamefont {Kelly}}]{Woon2005}%
  \BibitemOpen
  \bibfield  {author} {\bibinfo {author} {\bibfnamefont {K.~L.}\ \bibnamefont
  {Woon}}, \bibinfo {author} {\bibfnamefont {M.}~\bibnamefont {O’Neill}},
  \bibinfo {author} {\bibfnamefont {G.~J.}\ \bibnamefont {Richards}}, \bibinfo
  {author} {\bibfnamefont {M.~P.}\ \bibnamefont {Aldred}}, \ and\ \bibinfo
  {author} {\bibfnamefont {S.~M.}\ \bibnamefont {Kelly}},\ }
  \bibfield {title} {Stokes parameter studies of spontaneous emission from chiral nematic liquid crystals as a one-dimensional photonic stopband crystal: Experiment and theory}, \
  \href
  {http://link.aps.org/doi/10.1103/PhysRevE.71.041706} {\bibfield  {journal}
  {\bibinfo  {journal} {Physical Review E}\ }\textbf {\bibinfo {volume} {71}},\
  \bibinfo {pages} {41706} (\bibinfo {year} {2005})}\BibitemShut {NoStop}%
\bibitem [{\citenamefont {Ren}\ \emph {et~al.}(2012)\citenamefont {Ren},
  \citenamefont {Lu}, \citenamefont {Tan}, \citenamefont {Wu}, \citenamefont
  {Sun}, \citenamefont {Zhou}, \citenamefont {Xie}, \citenamefont {Sun},
  \citenamefont {Zhu}, \citenamefont {Jagadish}, \citenamefont {Shen},\ and\
  \citenamefont {Chen}}]{Ren2012}%
  \BibitemOpen
  \bibfield  {author} {\bibinfo {author} {\bibfnamefont {Q.}~\bibnamefont
  {Ren}}, \bibinfo {author} {\bibfnamefont {J.}~\bibnamefont {Lu}}, \bibinfo
  {author} {\bibfnamefont {H.~H.}\ \bibnamefont {Tan}}, \bibinfo {author}
  {\bibfnamefont {S.}~\bibnamefont {Wu}}, \bibinfo {author} {\bibfnamefont
  {L.}~\bibnamefont {Sun}}, \bibinfo {author} {\bibfnamefont {W.}~\bibnamefont
  {Zhou}}, \bibinfo {author} {\bibfnamefont {W.}~\bibnamefont {Xie}}, \bibinfo
  {author} {\bibfnamefont {Z.}~\bibnamefont {Sun}}, \bibinfo {author}
  {\bibfnamefont {Y.}~\bibnamefont {Zhu}}, \bibinfo {author} {\bibfnamefont
  {C.}~\bibnamefont {Jagadish}}, \bibinfo {author} {\bibfnamefont {S.~C.}\
  \bibnamefont {Shen}}, \ and\ \bibinfo {author} {\bibfnamefont
  {Z.}~\bibnamefont {Chen}},\ }
  \bibfield {title} {Spin-resolved Purcell effect in a quantum dot microcavity system}, \
  \href {\doibase 10.1021/nl3008083} {\bibfield
  {journal} {\bibinfo  {journal} {Nano letters}\ }\textbf {\bibinfo {volume}
  {12}},\ \bibinfo {pages} {3455} (\bibinfo {year} {2012})}\BibitemShut
  {NoStop}%
\bibitem [{\citenamefont {Konishi}\ \emph {et~al.}(2011)\citenamefont
  {Konishi}, \citenamefont {Nomura}, \citenamefont {Kumagai}, \citenamefont
  {Iwamoto}, \citenamefont {Arakawa},\ and\ \citenamefont
  {Kuwata-Gonokami}}]{Konishi2011}%
  \BibitemOpen
  \bibfield  {author} {\bibinfo {author} {\bibfnamefont {K.}~\bibnamefont
  {Konishi}}, \bibinfo {author} {\bibfnamefont {M.}~\bibnamefont {Nomura}},
  \bibinfo {author} {\bibfnamefont {N.}~\bibnamefont {Kumagai}}, \bibinfo
  {author} {\bibfnamefont {S.}~\bibnamefont {Iwamoto}}, \bibinfo {author}
  {\bibfnamefont {Y.}~\bibnamefont {Arakawa}}, \ and\ \bibinfo {author}
  {\bibfnamefont {M.}~\bibnamefont {Kuwata-Gonokami}},\ }
  \bibfield {title} {Circularly Polarized Light Emission from Semiconductor Planar Chiral Nanostructures}, \
  \href {\doibase
  10.1103/PhysRevLett.106.057402} {\bibfield  {journal} {\bibinfo  {journal}
  {Physical Review Letters}\ }\textbf {\bibinfo {volume} {106}},\ \bibinfo
  {pages} {057402} (\bibinfo {year} {2011})}\BibitemShut {NoStop}%
\bibitem [{\citenamefont {Maksimov}\ \emph {et~al.}(2014)\citenamefont
  {Maksimov}, \citenamefont {Tartakovskii}, \citenamefont {Filatov},
  \citenamefont {Lobanov}, \citenamefont {Gippius}, \citenamefont {Tikhodeev},
  \citenamefont {Schneider}, \citenamefont {Kamp}, \citenamefont {Maier},
  \citenamefont {H\"ofling},\ and\ \citenamefont {Kulakovskii}}]{Maksimov2014}%
  \BibitemOpen
  \bibfield  {author} {\bibinfo {author} {\bibfnamefont {A.~A.}\ \bibnamefont
  {Maksimov}}, \bibinfo {author} {\bibfnamefont {I.~I.}\ \bibnamefont
  {Tartakovskii}}, \bibinfo {author} {\bibfnamefont {E.~V.}\ \bibnamefont
  {Filatov}}, \bibinfo {author} {\bibfnamefont {S.~V.}\ \bibnamefont
  {Lobanov}}, \bibinfo {author} {\bibfnamefont {N.~A.}\ \bibnamefont
  {Gippius}}, \bibinfo {author} {\bibfnamefont {S.~G.}\ \bibnamefont
  {Tikhodeev}}, \bibinfo {author} {\bibfnamefont {C.}~\bibnamefont
  {Schneider}}, \bibinfo {author} {\bibfnamefont {M.}~\bibnamefont {Kamp}},
  \bibinfo {author} {\bibfnamefont {S.}~\bibnamefont {Maier}}, \bibinfo
  {author} {\bibfnamefont {S.}~\bibnamefont {H\"ofling}}, \ and\ \bibinfo
  {author} {\bibfnamefont {V.~D.}\ \bibnamefont {Kulakovskii}},\ }
  \bibfield {title} {Circularly polarized light emission from chiral spatially-structured planar semiconductor microcavities}, \
  \href
  {\doibase 10.1103/PhysRevB.89.045316} {\bibfield  {journal} {\bibinfo
  {journal} {Phys. Rev. B}\ }\textbf {\bibinfo {volume} {89}},\ \bibinfo
  {pages} {045316} (\bibinfo {year} {2014})}\BibitemShut {NoStop}%
\bibitem [{\citenamefont {Purcell}(1946)}]{Purcell1946}%
  \BibitemOpen
  \bibfield  {author} {\bibinfo {author} {\bibfnamefont {E.~M.}\ \bibnamefont
  {Purcell}},\ }
  \bibfield {title} {Spontaneous Emission Probabilities at Radio Frequencies}, \
  \href@noop {} {\bibfield  {journal} {\bibinfo  {journal}
  {Psysical Review}\ }\textbf {\bibinfo {volume} {69}},\ \bibinfo {pages} {681}
  (\bibinfo {year} {1946})}\BibitemShut {NoStop}%
\bibitem [{\citenamefont {Bykov}(1972)}]{Bykov1972}%
  \BibitemOpen
  \bibfield  {author} {\bibinfo {author} {\bibfnamefont {V.~P.}\ \bibnamefont
  {Bykov}},\ }
  \bibfield {title} {Spontaneous Emission in a Periodic Structure}, \
  \href@noop {} {\bibfield  {journal} {\bibinfo  {journal} {JETP}\
  }\textbf {\bibinfo {volume} {35}},\ \bibinfo {pages} {269} (\bibinfo {year}
  {1972})}\BibitemShut {NoStop}%
\bibitem [{\citenamefont {Tikhodeev}\ \emph {et~al.}(2002)\citenamefont
  {Tikhodeev}, \citenamefont {Yablonskii}, \citenamefont {Muljarov},
  \citenamefont {Gippius},\ and\ \citenamefont {Ishihara}}]{Tikhodeev2002}%
  \BibitemOpen
  \bibfield  {author} {\bibinfo {author} {\bibfnamefont {S.~G.}\ \bibnamefont
  {Tikhodeev}}, \bibinfo {author} {\bibfnamefont {A.~L.}\ \bibnamefont
  {Yablonskii}}, \bibinfo {author} {\bibfnamefont {E.~A.}\ \bibnamefont
  {Muljarov}}, \bibinfo {author} {\bibfnamefont {N.~A.}\ \bibnamefont
  {Gippius}}, \ and\ \bibinfo {author} {\bibfnamefont {T.}~\bibnamefont
  {Ishihara}},\ }
  \bibfield {title} {Quasiguided modes and optical properties of photonic crystal slabs}, \
  \href {\doibase 10.1103/PhysRevB.66.045102} {\bibfield
  {journal} {\bibinfo  {journal} {Physical Review B}\ }\textbf {\bibinfo
  {volume} {66}},\ \bibinfo {pages} {045102} (\bibinfo {year}
  {2002})}\BibitemShut {NoStop}%
\bibitem [{\citenamefont {Whittaker}\ and\ \citenamefont
  {Culshaw}(1999)}]{Whittaker1999}%
  \BibitemOpen
  \bibfield  {author} {\bibinfo {author} {\bibfnamefont {D.~M.}\ \bibnamefont
  {Whittaker}}\ and\ \bibinfo {author} {\bibfnamefont {I.~S.}\ \bibnamefont
  {Culshaw}},\ }
  \bibfield {title} {Scattering-matrix treatment of patterned multilayer photonic structures}, \
  \href {\doibase 10.1103/PhysRevB.60.2610} {\bibfield  {journal}
  {\bibinfo  {journal} {Physical Review B}\ }\textbf {\bibinfo {volume} {60}},\
  \bibinfo {pages} {2610} (\bibinfo {year} {1999})}\BibitemShut {NoStop}%
\bibitem [{\citenamefont {Weiss}\ \emph {et~al.}(2009)\citenamefont {Weiss},
  \citenamefont {Granet}, \citenamefont {Gippius}, \citenamefont {Tikhodeev},\
  and\ \citenamefont {Giessen}}]{Weiss2009}%
  \BibitemOpen
  \bibfield  {author} {\bibinfo {author} {\bibfnamefont {T.}~\bibnamefont
  {Weiss}}, \bibinfo {author} {\bibfnamefont {G.}~\bibnamefont {Granet}},
  \bibinfo {author} {\bibfnamefont {N.~A.}\ \bibnamefont {Gippius}}, \bibinfo
  {author} {\bibfnamefont {S.~G.}\ \bibnamefont {Tikhodeev}}, \ and\ \bibinfo
  {author} {\bibfnamefont {H.}~\bibnamefont {Giessen}},\ }
  \bibfield {title} {Matched coordinates and adaptive spatial resolution in the Fourier modal method}, \
  \href
  {http://www.ncbi.nlm.nih.gov/pubmed/21293528} {\bibfield  {journal} {\bibinfo
   {journal} {Optics express}\ }\textbf {\bibinfo {volume} {17}},\ \bibinfo
  {pages} {8051} (\bibinfo {year} {2009})}\BibitemShut {NoStop}%
\bibitem [{\citenamefont {Lobanov}\ \emph {et~al.}(2012)\citenamefont
  {Lobanov}, \citenamefont {Weiss}, \citenamefont {Dregely}, \citenamefont
  {Giessen}, \citenamefont {Gippius},\ and\ \citenamefont
  {Tikhodeev}}]{Lobanov2012}%
  \BibitemOpen
  \bibfield  {author} {\bibinfo {author} {\bibfnamefont {S.~V.}\ \bibnamefont
  {Lobanov}}, \bibinfo {author} {\bibfnamefont {T.}~\bibnamefont {Weiss}},
  \bibinfo {author} {\bibfnamefont {D.}~\bibnamefont {Dregely}}, \bibinfo
  {author} {\bibfnamefont {H.}~\bibnamefont {Giessen}}, \bibinfo {author}
  {\bibfnamefont {N.~A.}\ \bibnamefont {Gippius}}, \ and\ \bibinfo {author}
  {\bibfnamefont {S.~G.}\ \bibnamefont {Tikhodeev}},\ } %
  \bibfield {title} {Emission properties of an oscillating point dipole from a gold
  Yagi-Uda nanoantenna array}, \
  \href {\doibase
  10.1103/PhysRevB.85.155137} {\bibfield  {journal} {\bibinfo  {journal}
  {Physical Review B}\ }\textbf {\bibinfo {volume} {85}},\ \bibinfo {pages}
  {155137} (\bibinfo {year} {2012})}\BibitemShut {NoStop}%
\end{thebibliography}
\end{document}